\documentclass{article}
\begin{document}
\title{RED SHIFT FROM GRAVITATIONAL BACK REACTION}
\author{Ernst Fischer\footnote{Auf der H\"ohe 82, D-52223 Stolberg, Germany}}
\date{e.fischer.stolberg@t-online.de}
\maketitle

\begin{abstract}
Deviations from geodesic motion caused by gravitational radiation have been
discussed in the last decades to describe the motion of particles or photons
in strong fields around collapsed objects. On cosmological scale this effect,
which in the first order is caused by the finite speed of gravitational
interaction, is important also in the weak field limit. In this paper the
energy loss by transfer to the gravitational potential is determined in a
quasi-Newtonian approximation for the examples of a static Einstein universe
and for an expanding universe with flat metric. In both cases the resulting
red shift is a considerable fraction of the total red shift and requires an
adjustment of the age and the matter composition in our models of the
universe.
\end{abstract}

\section{Introduction}
Today the theory of general relativity (GRT) is accepted as the correct
description of gravitation, but due to its non-linear character and the
complicated mathematical formalism by now the application to practical
problems is restricted to approximate solutions in most cases. Especially in
the weak field limit most of the work, beginning with the 'classical'
problems discussed by Einstein, like perihelion advance or gravitational
aberration, is restricted to the 'geodesic approximation'. That means that
the trajectories of test masses or photons are determined as geodesic motions
in non-Minkowskian space-time with the metric set up by the surrounding
matter fields. A comprehensive description of this method can be found in
every textbook on GRT (see e.g. R.M.Wald 1984).

But of course, distinction between field masses and test masses is somewhat
artificial, as according to GRT every matter particle or energy quantum
contributes to the metric. Thus by principle any motion of a test particle or
photon changes the metric in its neighbourhood, which then leads to a back
reaction on the particle itself. In the last decades with the advent of
growing observational information on compact sources of gravitation like
neutron stars and black holes the need has increased for a better description
of deviation from geodesic motion in curved space, caused by back reaction
processes. Not only the loss of angular momentum by emission of gravitational
radiation, observed in the rotational motion of double pulsars (Weisberg and
Tailor 1984), but also the inspiralling of matter in the accretion disk of
black holes have focussed increasing attention to the problem of back
reaction. A review of the state of activities in this field has been given by
Poisson (Poisson 2004).

Due to the non-linear form of the basic equations of GRT analytical solutions
require an enormous effort of mathematical calculations, even if deviations
from geodesic motion is treated as small perturbations. But common to all the
papers in the field is the result that the deviation from geodesic motion in
curved space the leading order of the perturbation results in an energy loss,
which is emitted as gravitational radiation. It is this gravitational
radiation, which attracts the interest of many researchers, as they hope to
prove the existence of this radiation experimentally on earth. Of course,
these people are looking for events which lead to strong radiation pulses, as
they are expected from situations near the surface of black holes.

These problems may require higher order perturbation models, but the effect
of energy loss by gravitational radiation is not limited to the strong field
regime, but affects all motions in mass filled space. By principle every
moving particle or photon will suffer such an energy loss, even if the effect
is so tiny that we never can expect to observe it experimentally. But we
cannot exclude that by accumulation over cosmological times the energy loss
may be important. If it plays a role, this would lead to corrections of our
cosmological models.

\section{Energy loss in curved space}
To estimate the order of magnitude of the expected effects we must not go
into the full treatment by GRT. The more exact calculations mentioned above
have shown that in the leading order the deviations from geodesic motion are
caused by the fact that gravitational interaction is limited to the speed of
light. The most important difference is, compared to Newtonian gravity, that
we have to use retarded forces to solve the equation of motion. It is this
retarded interaction which gives rise to the loss of energy by transfer to
the gravitational potential. Thus, to calculate this effect in first order,
we will study the behaviour of a test particle, moving with respect to a
reference frame, given by a homogeneous mass-filled universe. We will study
the effects in a quasi-Newtonian approximation, but with the constraints
imposed by general relativity:
\begin{enumerate}
\item Gravitational interaction is limited by the speed of light.

\item Mass or energy cause an intrinsic curvature of space. A
spatially homogeneous universe can thus be regarded as a three-dimensional
surface of constant curvature. By restriction to closed space we avoid the
infinity problems of Newtonian physics.

\item Lines of force between masses follow the geodesic lines connecting them.
This mixture of the geometrical description of GRT with the force field
description on Newtonian theory appears justified, as we restrict our
calculations to the weak field limit.
\end{enumerate}
To avoid confusion with effects stemming from time-dependence of the metric,
we first confine the discussion to the time-independent metric of a static
Einstein universe. To describe the motion on a surface of constant curvature,
it is convenient to use 4-dimensional spherical coordinates defined by
\begin{eqnarray}
\label{coord}
x=r \cos \gamma \cos \theta \cos \phi &,&y=r \cos \gamma \cos \theta \sin
\phi ,\nonumber \\z=r \cos \gamma \sin \theta &, &w= r \sin \gamma
\end{eqnarray}
where $x,y,z$ and $w$ is a set of four Cartesian coordinates. In this system
we determine the force exerted on a test mass at $P=(R,0,0,0)$ by the mass
contained in a volume element $dV$ of the surface $r=R$, the volume element
being given by
\begin{equation}
\label{vol}
dV= R^3 \cos ^2 \gamma \cos \theta \:d\gamma \:d\theta \:d\phi
\end{equation}
The distance between the element and $P$ as measured along the geodesic of
the surface $r=R$ is given by
\begin{equation}
\label{s}
s= 2R \arcsin \frac {r_P}{2R}=2R \arcsin \sqrt{\frac{1-\cos \gamma \cos
\theta \cos \phi}{2}}
\end{equation}
The component of gravitational force at point $P$ in some direction, defined
by the unit vector $\vec e$, is given by
\begin{equation}
\label{dF}
dF=\frac{G\rho m dV}{s^2} (\vec{e_s} \vec e ),
\end{equation}
where $\rho$ is the mass density, $G$ the gravitational constant and $\vec
{e_s}$ the unit vector in the direction of the geodesic at point $P$. As all
directions are equivalent, we can choose $\vec e$ in the direction of the
y-coordinate. In this case the projection onto the direction of the force
component is
\begin{equation}
\label{ee}
(\vec{e_s} \vec e )= \frac {\cos \gamma \cos\theta \sin \phi}{ \sqrt{1-\cos
^2\gamma \cos ^2 \theta \cos ^2 \phi}}
\end{equation}
For a universe with constant mass density we obtain the total force in the
y-direction by integrating equation (\ref{dF}) over all distances and
directions:
\begin{equation}
\label{Fy}
F_y = G \rho m R^3 \int _{-\infty} ^{\infty}\int _{-\pi /2}^{\pi /2}
\int _{-\pi /2}^{\pi /2}
 \frac {\cos ^3\gamma  \cos ^2\theta  \sin \phi}{ s^2 \sqrt{1-\cos ^2
\gamma\cos ^2 \theta \cos ^2 \phi}} \,d\gamma \,d\theta \,d\phi .
\end{equation}
The limits of integration over the angle $\phi$ are set to $\pm \infty$, as
the lines of force may extent beyond the reciprocal pole $(-R,0,0,0)$.

For any mass at rest with respect to the frame of reference it is obvious
that this integral is zero due to symmetry. In the Newtonian limit this holds
also for a moving mass, as the gravitational force from matter in all volume
elements is assumed to act instantaneously. If, however, the velocity of
interaction is limited to the speed of light, we have to use retarded
coordinates. Thus to determine the force at some instant $t$ we have to
replace the distance $s(t)$ by $s'(t) = s(t-\tau)$, with $\tau$ being the
running time of the signal $\tau = s'/c$. That means that the integral does
not remain symmetrical. All distances in the direction of motion are enhanced
and all distances in the direction opposite to the motion are reduced. As a
result we find that there exists a force that tends to reduce the momentum of
every mass or energy quantum moving relative to the rest frame of the
universe. Due to the finite velocity of interaction gravity acts like a kind
of viscosity of mass-filled space, that slows down every motion in the
universe and transfers energy to the gravitational potential. It is this
transfer to the gravitational potential, which is commonly labeled as
gravitational radiation.

For the numerical evaluation of the force integral it is convenient to change
the coordinate system. Instead of replacing the distance $s$ and its
projection onto the direction of motion in equation (\ref{Fy}), we keep these
quantities unchanged and change the integration variable instead, introducing
the co-moving angular coordinate
\begin{equation}
\label{phi}
\phi '= \phi (t')= \phi (t)+ \frac{d\phi}{dt} \tau =  \phi (t)+ \frac{vs'}{Rc}
\end{equation}
where $v$ is the velocity of the moving mass. This can be done, as $\phi$ is
the only coordinate which is affected by the motion. Expressing the distance
$s'$ according to equation (\ref{s}), we find the differential
\begin{equation}
\label{dphi}
d\phi = d\phi '\left ( 1-\frac{v}{c}\frac {\cos \gamma \cos\theta \sin
\phi '}{\sqrt{1-\cos ^2\gamma \cos ^2 \theta \cos ^2 \phi '}}\right)
\end{equation}
The limits of the integral are not changed by the transformation. So finally,
omitting the primes for convenience, for the total force we obtain the
expression
\begin{eqnarray}
\label{Fy2}
F_y &=& -2G \rho m R^3 \frac {v}{c}\int _{-\infty} ^{\infty}\int _{-\pi
/2}^{\pi /2}
\int _{-\pi /2}^{\pi /2} \times \nonumber\\ &\times&
 \frac {\cos ^4  \gamma\cos ^3\theta \sin^2 \phi \:d\gamma \:d\theta \:d\phi }{ (1-\cos ^2
\gamma \cos ^2\theta \cos ^2 \phi)\left(\arcsin\sqrt{(1-\cos \gamma \cos \theta
\cos \phi)/2}\right)^2}
\end{eqnarray}
The integral has been solved numerically. Its value is $Y=3.0695$

As the resulting force is always in line with the momentum vector $\vec p = m
\vec v$, we can rewrite equation (\ref{Fy2}) in vectorial form, replacing $F$
by the time derivative of momentum
\begin{equation}
\label{dp}
\frac{d \vec p}{dt}=-\frac{2Y \rho GR}{c} \vec p.
\end{equation}
Accordingly the loss of kinetic energy is given by
\begin{equation}
\label{dE}
\frac{dE}{dt}=\frac{d}{dt}\frac{p^2}{2m}=-\frac{4Y \rho GR}{c} E.
\end{equation}
As these equations have been derived without any restriction of the particle
velocity, they should be valid also for relativistic motion and thus also for
photons. As in this case the velocity is fixed to the speed of light, the
energy loss will show up as a change of frequency or wave length:
\begin{equation}
\label{lambda}
\lambda =\lambda _0 \exp \left( \frac{4Y \rho GR}{c}t \right ),
\end{equation}
implying, as a first approximation, a red shift increasing linearly with the
distance of the source. In an expanding universe this red shift would be
superimposed to the red shift resulting from the change of metric. The
quantity
\begin{equation}
\label{HR}
 H_R =4Y \rho GR/c
\end{equation}
must be regarded as an additional part of the observed Hubble constant H. To
estimate the contribution of $H_R$ on $H$, we consider a universe at the
critical closing density $\rho_{cr}=3H^2/(8\pi G)$. The radius of the
corresponding Einstein universe is given by $R=\sqrt{c^2/(4\pi G\rho)}$.
Introducing these values into Eq.(\ref{HR}) we find the relation
\begin{equation}
\label{HRH}
 H_R =\sqrt{\frac{3}{2}}\,\frac{Y}{\pi}\times H = 1.197 H .
\end{equation}
With other words: the red shift caused by energy loss due to gravitational
radiation could account for the complete observed red shift, leaving no room
for expansion at all. Exact agreement would be obtained at $\rho = 0.7
\rho_{cr}$. Of course, a quantitative comparison of the effect in a static
Einstein universe with that in an expanding universe in this way is not
correct. But it is obvious that an analogous energy loss must be present in
this case, too.

\section{Energy loss in expanding flat space}
While the Einstein universe is curved in a definite way, the curvature of an
expanding universe may be much lower or even zero. But also in this case
particles and photons move through mass filled space and thus lose energy and
feel the "gravitational viscosity".

The energy loss rate can be easily estimated for the limiting case of an
expanding universe with flat geometry. In this case space is unlimited, but
gravitational interaction is limited to the region which is causally
connected to the moving particle. Denoting the fraction of the Hubble
constant due to expansion by $H_E$, interaction is limited to a sphere of
radius $R=c/H_E$.

To determine the radiative energy loss rate, we consider the motion of a
particle in the centre of a mass filled sphere of Radius R, moving in
x-direction of a Cartesian coordinate system. The contribution of a volume
element at distance $\vec s$ to the x-component of the gravitational force is
\begin{equation}
dF=\frac{G\rho m dV}{s^2} (\vec{e_s} \vec {e_x} ).
\end{equation}
Analogous to the case of the Einstein universe, to include retardation of
interaction, we have to introduce retarded distances or, changing to comoving
coordinates, instead of $x'$ we have to use the comoving coordinate
\begin{equation}
x= x'-\frac{vs'}{c} \quad \rm{and} \quad dx= dx'
\left(1-\frac{vx'}{s'c}\right)
\end{equation}
Using spherical coordinates, integration of the force over a sphere of radius
$R$ leads to
\begin{equation}
F_x = -G \rho m \frac{v}{c}\int _{0} ^{R} dr \int _{-\pi /2}^{\pi /2}\cos ^3
\theta \,d\theta \int _{0} ^{2\pi}\cos ^2 .
\phi\,d\phi.
\end{equation}
From this equation we find $H_R$ in a similar way as for the Einstein
universe. Using $R=c/H_E$ and the value for the density $\rho=3H_E^2/(8\pi
G)$, which follows from the GRT field equations in the case of a flat metric,
we get
\begin{equation}
 H_R =\frac{8\pi}{3}\, \frac{\rho G}{c}\,R =\frac{8\pi}{3}\,\frac{3H_E^2}{8\pi G}
 \,\frac{G}{c}\,\frac{c}{H_E}\,=\,H_E
\end{equation}
That means that only half of the observed red shift can be attributed to
expansion, the other half is caused by energy loss to the gravitational
potential. Thus, compared to the presently favoured standard model, the age
of the universe is doubled. This may explain the fact that we see fully
developed galaxies with metallicities comparable to nearby ones at red shifts
up to z=6. Also the number of absorption lines, the Lyman forest, in the
spectra of distant quasars implies a much higher age than discussed in the
'concordance model' (see Liebscher, Priester, Hoell 1992). As the mean
density according to the GRT field equation scales with the square of $H_E$,
the total density is reduced to 25\% of the presently discussed values,
making the adoption of a dark energy component unnecessary, at least from the
viewpoint of red shift.

Though the results presented here may be inaccurate due to unjustified
simplifications, the fact remains that, if we trust GRT, the energy loss of
moving particles and energy quanta exists, and it is important not only in
the range of strong gravitational fields, but also for the global processes
which determine the history of the universe.


\begin{thebibliography}{}
\bibitem{wald} R. M. Wald, {\it General Relativity}, (Univ. of Chicago Press, Chicago, USA, 1984)
\bibitem{weisberg} I. M. Weisberg, and I. H. Taylor, Phys. Rev. Lett. 52, 1348 (1984)
\bibitem{poisson}  E. Poisson, {\it The motion of point particles in curved spacetime},
Living Rev. Relativity 7 (2004) and (arXiv gr-qc/0306052)
\bibitem{liebscher} D. E. Liebscher, W. Priester, J. Hoell, Astron. Nachr. 313, 265-273 (1992)
\end{thebibliography}
\end{document}